\pdfoutput=1
\documentclass[]{smule}
\usepackage{amsmath}
\usepackage{amssymb}
\usepackage{array}
\usepackage{libertinus}

\title{Convolutional VocalTrax: Replacing Backpropagation Through Time with Convolution in Differentiable Articulatory Synthesis}

\author[1\dagger]{Yongyi Zang}
\author[2\dagger]{Debasish Ray Mohapatra}
\author[1]{Randal Leistikow}
\author[1]{Perry R. Cook}

\affiliation[1]{Smule Labs}
\affiliation[2]{University of British Columbia, Canada}

\contribution[\dagger]{These authors contributed equally.}

\abstract{
VocalTrax recovers interpretable articulatory controls from a recording by gradient descent through a differentiable Pink Trombone synthesizer. Its vocal tract is a Kelly-Lochbaum waveguide updated twice per audio sample, so every gradient step backpropagates through a long recurrence, which is slow on GPUs. We propose Convolutional VocalTrax, which computes the exact frequency response of the same waveguide and applies it as a causal convolution. We assume that the tract is stationary within 5.8~ms blocks and that its lossy impulse response can be truncated after 93~ms. Gradients then pass through FFTs and a short recursion along the tract instead of through time. On 600 AudioMNIST recordings, the tract forward and backward pass runs 47 to 63 times faster on an A100 GPU, and a classifier trained only on real speech recognizes our reconstructions more often than those of VocalTrax (14.5\% vs.\ 10.0\%), although the log-mel distance to the target is 11\% higher. Parameters fitted with our model keep this gain when rendered by the original VocalTrax (13.8\%).
}

\metadata[Keywords]{Articulatory synthesis, acoustic-to-articulatory inversion, differentiable signal processing, vocal tract, waveguide}

\begin{document}

\maketitle

\section{Introduction}
\label{sec:intro}

Articulatory synthesis generates speech by simulating the vocal apparatus, which makes its controls interpretable and easy to edit \citep{birkholzModelingConsonantVowelCoarticulation2013}. Reproducing a given recording with such a synthesizer requires acoustic-to-articulatory inversion (AAI). Data-driven AAI learns this mapping from parallel acoustic and articulatory measurements \citep{dangEstimationVocalTract2002,liuDeepRecurrentApproach2015}, which are expensive to collect. Analysis-by-synthesis instead searches for the controls that make the synthesizer reproduce the target \citep{atalInversionArticulatoryAcoustic1978,richmondEstimatingArticulatoryParameters2001}. Derivative-free search has been used for this purpose \citep{schleusingJointSourceFilterOptimization2013,gaoArticulatoryCopySynthesis2019,camara2023optimization}, but it needs many evaluations of the synthesizer.

Differentiable implementations allow analysis-by-synthesis with gradient descent \citep{engelDDSPDifferentiableDigital2020,kuznetsov2020differentiable}. \citet{sudholt2023vtg} estimate vowel area functions for the Pink Trombone \citep{pt} by matching the magnitude of its analytical transfer function. VocalTrax \citep{mo2024vocaltrax} extends gradient-based matching to connected speech and other vocalizations. It implements the Pink Trombone in JAX \citep{bradbury2018jax}, estimates the fundamental frequency with CREPE \citep{kim2018crepe}, and optimizes frame-wise tongue, constriction and tenseness controls end-to-end with a log-mel loss. It needs no articulatory data, and its reconstructions of real speech are recognized more accurately than those of earlier gradient-based estimation \citep{mo2024vocaltrax}.

Fitting VocalTrax is slow, however. Its vocal tract is a Kelly-Lochbaum waveguide \citep{kellySpeechSynthesis1962} that is advanced twice per audio sample, so automatic differentiation backpropagates through about 88,000 sequential updates for every second of audio. Backpropagation through long recurrences is expensive and, in general, prone to vanishing or exploding gradients \citep{pascanu2013difficulty}. On an A100 GPU, the backward pass through the tract alone takes 0.74~s per second of audio, and a 1000-step fit of a one-second clip takes about 24 minutes.

In this paper we propose Convolutional VocalTrax, which removes the time recursion from the tract and from its gradient. We rely on two assumptions that are common in speech acoustics: the tract geometry changes little within a few milliseconds, and the losses in the tract make its impulse response decay within tens of milliseconds. Under these assumptions, each short block of the glottal source can be filtered with a fixed impulse response, which we compute exactly from the waveguide in the frequency domain. Our contributions are as follows. (1) We derive the frequency response of the waveguide as it is executed in VocalTrax, including its half-sample updates, and show that it reduces to a single evaluation when the number of segments is even. For fixed geometry, the resulting convolution reconstructs the output of VocalTrax exactly, up to a truncation of the impulse response that we show to be negligible. (2) We use this response in a frozen-block convolution whose gradients pass through FFTs and a 43-step recursion along the tract. This makes the tract forward and backward pass 47 to 63 times faster on a GPU. (3) On 600 AudioMNIST recordings \citep{AudioMNIST}, a classifier trained only on real speech recognizes our reconstructions more accurately than those of VocalTrax, and the fitted parameters keep this advantage when they are rendered by the original VocalTrax.

\section{Related Work}
\label{sec:related}

\textbf{Articulatory inversion and synthesis.} Neural networks can map between articulatory and acoustic representations \citep{sahaLearningJointArticulatoryacoustic2020,liuDeepRecurrentApproach2015}, but they need paired articulatory and acoustic data. Physics-based synthesizers such as VocalTractLab avoid this requirement and have been compared with text-to-speech systems in intelligibility and naturalness \citep{krugIntelligibilityNaturalnessArticulatory2021}. The Pink Trombone \citep{pt} has a simpler tract of 44 cylindrical segments and has been used in several sound matching studies \citep{sudholt2023vtg,camara2023optimization,mo2024vocaltrax}. The method of \citet{sudholt2023vtg} fits the magnitude of the analytical transfer function of this tract \citep{smythTransferFunctionPiecewiseCylindrical2021} to a vocal tract filter obtained by inverse filtering, one frame at a time, and synthesizes audio only after the fit. The recurrent waveguide of VocalTrax has recently been extended with a sublingual second source, a B-spline tract shape and learnable damping to copy-synthesize biphonic singing \citep{camara2026biphonic}, and it is still differentiated through the sample-by-sample recursion. We also use the transfer function, but we keep its phase and place it inside the end-to-end time-domain fit of VocalTrax, where the geometry moves from frame to frame.

\textbf{Fast simulation and differentiable filters.} Two-dimensional finite-difference models reduce the cost of vocal tract acoustics compared with full three-dimensional simulation \citep{zappi2016towards,mohapatra2019extended}. Physics-informed neural networks have recently been trained to solve the vocal tract equations \citep{yokota2026physics,lu2026learning}, which requires training for each set of conditions. In differentiable signal processing, recursive filters have been trained both in the time domain and through their frequency response \citep{kuznetsov2020differentiable,colonelDirectDesignBiquad2022}, and time-varying filters are commonly applied frame by frame with FFT convolution \citep{engelDDSPDifferentiableDigital2020}. We apply the latter idea to a physical waveguide whose frequency response we derive exactly.

\section{Method}
\label{sec:method}

\subsection{VocalTrax}
\label{ssec:vocaltrax}

VocalTrax synthesizes $\hat{y}=V(u;\theta)$ from a glottal source $u=G(F_0,T)$ \citep{mo2024vocaltrax}. The source is the Liljencrants-Fant model of the glottal flow derivative \citep{fantFourparameterModelGlottal1985}, driven by a CREPE estimate of $F_0$ and a frame-wise tenseness $T$, with aspiration noise proportional to $1-\sqrt{T}$. The tract has $S=44$ segments whose diameters are set every 1024 samples by the tongue position and diameter and by a throat and a lip constriction. With segment areas $A_i$, the reflection coefficients are
\begin{equation}
r_i=\frac{A_i-A_{i+1}}{A_i+A_{i+1}}.
\label{eq:refl}
\end{equation}
Right- and left-going waves $R_i$ and $L_i$ meet at each junction, where $w_i=r_i(R_i+L_{i+1})$ gives the outgoing waves $R_i-w_i$ and $L_{i+1}+w_i$. Every propagation step scales the waves by $d=0.999$, and the glottis and lips reflect with $r_g=0.75$ and $r_l=-0.85$. The waveguide runs two half-sample updates per audio sample. The source enters on the first update, and the output is a quarter of the sum of the two waves arriving at the lips. The reflection coefficients are interpolated to every sample, and the loss is the mean squared error between log-mel spectrograms. JAX differentiates the loop over samples, so the gradient for a clip of $N$ samples passes through $2N$ sequential updates.

\subsection{Exact stationary response}
\label{ssec:response}

For fixed reflection coefficients, the junctions can be eliminated in the frequency domain. At the half-sample clock, let $a(\omega)=d\,e^{-j\omega}$ and $q_i=L_i/R_i$. Starting from $q_{S-1}=a\,r_l$ at the lips,
\begin{equation}
q_i=a\,\frac{r_i+a\,q_{i+1}}{1+a\,r_i\,q_{i+1}},\qquad
t_i=\frac{R_{i+1}}{R_i}=a\,\frac{1-r_i}{1+a\,r_i\,q_{i+1}},
\label{eq:recursion}
\end{equation}
and the response from the injected source to the wave at the lips is
\begin{equation}
H_h(\omega)=\frac{a}{1-a\,r_g\,q_0}\prod_{i=0}^{S-2}t_i .
\label{eq:halfrate}
\end{equation}
Since the source enters on alternate half steps and the two lip samples are summed, the response at the audio rate is
\begin{equation}
H(\omega)=\tfrac{1}{8}\Big[(1+e^{j\omega/2})H_h(\tfrac{\omega}{2})+(1-e^{j\omega/2})H_h(\tfrac{\omega}{2}+\pi)\Big].
\label{eq:polyphase}
\end{equation}
Replacing $\omega$ with $\omega+\pi$ changes the sign of $a$, $q_i$ and $t_i$, so $H_h(\omega+\pi)=(-1)^S H_h(\omega)$. For $S=44$ the second term in (\ref{eq:polyphase}) cancels and $H(\omega)=H_h(\omega/2)/4$. In other words, every path from the glottis to the lips takes an even number of half steps. Our implementation evaluates both terms of (\ref{eq:polyphase}). The impulse response of the recurrent waveguide, computed directly in double precision, matches (\ref{eq:polyphase}) to a relative error of $6\times10^{-10}$. The recursion (\ref{eq:recursion}) runs over 43 junctions and is vectorized over all frequencies.

\subsection{Frozen-block convolution}
\label{ssec:convolution}

For moving geometry we make two assumptions. First, the geometry is constant within blocks of $H=256$ samples (5.8~ms at 44.1~kHz), and for block $b$ we interpolate the reflection coefficients at its midpoint. Second, the response of the lossy tract decays, so we keep the first $L=4096$ samples (93~ms) of its impulse response $h_b$, obtained by an inverse FFT of (\ref{eq:polyphase}) on a grid of 8192 frequencies. The output is
\begin{equation}
\hat{y}[n]=\sum_{b}\sum_{m=0}^{H-1}u[bH+m]\,h_b[n-bH-m],
\label{eq:conv}
\end{equation}
which we compute with zero-padded FFTs and overlap-add, so the convolution is linear rather than circular. The source, controls, loss and optimizer are unchanged. Gradients with respect to the controls pass through the FFTs and the recursion (\ref{eq:recursion}), and all blocks and frequencies are processed in parallel. The length of this gradient path does not depend on the duration of the clip.

When the geometry is fixed, the waveguide is a linear time-invariant system, and (\ref{eq:conv}) reconstructs the output of VocalTrax exactly except for the truncation of $h_b$ to $L$ samples. This truncation is negligible for a lossy tract. For the recurrent model, at most $1.2\times10^{-11}$ of the energy of the sensitivity of an output sample to the source lies more than 100~ms in the past, for neutral, constricted and fitted tracts alike (Table~\ref{tab:validation}). In double precision the two impulse responses agree to a relative error of $6\times10^{-10}$, and in single precision the waveforms agree to $8.1\times10^{-6}$, which is the level of round-off.

The approximation therefore enters only when the geometry moves. In (\ref{eq:conv}) the tail of each block keeps the filter of the block that excited it, while in the waveguide the waves already in the tract propagate through the current geometry. We measure this error on a noise-excited tract whose diameters vary sinusoidally across control frames by 5\% and 20\% of their mean. The error grows with the speed of the motion and with the block length (Table~\ref{tab:validation}). Blocks of $H=256$ samples keep it at 1.0\% and 4.0\% for these motions, at half the cost of 64-sample blocks.

\begin{table}[t]
\centering
\caption{Validation of the two assumptions. Top: relative waveform error of our model against VocalTrax for fixed geometry and for diameter motion of 5\% and 20\%, and CPU time of one forward pass over 8192 samples. Bottom: sensitivity of one output sample of VocalTrax to past source samples, as the lag containing 99\% and 99.9\% of the energy and the energy fraction beyond 100~ms.}
\label{tab:validation}
\setlength{\tabcolsep}{5pt}
\begin{tabular}{rcccc}
\toprule
Block $H$ & Fixed & 5\% motion & 20\% motion & Time (s) \\
\midrule
64 & $8.1\times10^{-6}$ & 0.79\% & 2.95\% & 0.36 \\
128 & $8.1\times10^{-6}$ & 0.86\% & 3.33\% & 0.19 \\
256 & $8.1\times10^{-6}$ & 1.04\% & 4.00\% & 0.17 \\
512 & $8.1\times10^{-6}$ & 1.46\% & 5.55\% & 0.11 \\
\bottomrule
\end{tabular}

\vspace{2mm}
\begin{tabular}{lccc}
\toprule
Tract geometry & 99\% lag & 99.9\% lag & Beyond 100~ms \\
\midrule
Neutral & 4.5 ms & 6.5 ms & $2.9\times10^{-35}$ \\
Constricted & 6.7 ms & 13.2 ms & $1.2\times10^{-11}$ \\
Fitted by VocalTrax & 16.3 ms & 37.7 ms & $3.2\times10^{-18}$ \\
Fitted by our model & 10.8 ms & 17.3 ms & $2.6\times10^{-18}$ \\
\bottomrule
\end{tabular}
\end{table}

\subsection{Parameter transfer}
\label{ssec:transfer}

To test whether the fitted parameters are meaningful for the original model, we render them with the recurrent VocalTrax, using the same $F_0$ and noise, without any further optimization.

\begin{figure}[t]
\centering
\includegraphics[width=\linewidth]{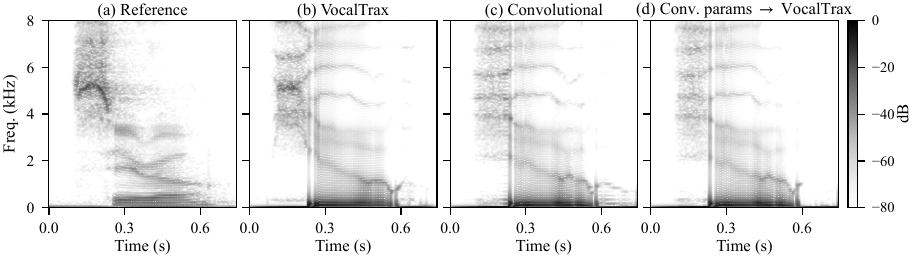}
\caption{Spectrograms of the digit ``zero'' (speaker 55) and its 1000-step reconstructions. (d) renders the parameters fitted by our model with the original recurrent VocalTrax, without further optimization. All panels share one dB scale, and no signal is normalized.}
\label{fig:spec}
\end{figure}

\section{Experiments}
\label{sec:experiments}

\subsection{Setup}
\label{ssec:setup}

We use 600 recordings from AudioMNIST \citep{AudioMNIST}: 12 held-out speakers, 10 digits and 5 repetitions per digit, between 0.42 and 1.0~s long. Each recording is fitted independently with the protocol in Table~\ref{tab:protocol}, shared by both models. It keeps the optimizer of VocalTrax \citep{mo2024vocaltrax}, namely AdamW \citep{adamw}, projection to $[0,1]$ and Savitzky-Golay smoothing \citep{savitzkySmoothingDifferentiationData1964}. We add constraints so that both models see identical inputs and stay within the bounds: resampled and padded targets, $F_0$ interpolated onto the control grid, a second projection after each smoothing because the smoother can overshoot, and the same aspiration noise for both models at every step.

\begin{table}[t]
\centering
\caption{Fitting protocol shared by both models.}
\label{tab:protocol}
\small
\begin{tabular}{@{}l>{\raggedright\arraybackslash}p{0.78\linewidth}@{}}
\toprule
Audio & Resampled to 44.1 kHz, right-padded to a multiple of 1024 samples, one extra control frame at the end \\
$F_0$ & CREPE full model, 23 ms hop; confidence $\le 0.5$ set to 1 Hz; interpolated to the control frames; fixed during fitting \\
Controls & 5 per 1024-sample frame, normalized to $[0,1]$: tongue diameter $[2.05, 3.5]$, tongue position $[12, 29]$, throat and lip constriction $[-0.99, 0.99]$, tenseness $[0.1, 1]$ \\
Init. & Tract controls 0.5, tenseness 1 \\
Optimizer & AdamW, learning rate 0.01, weight decay $10^{-4}$, 1000 steps \\
Box & Projection to $[0,1]$ after every update and every smoothing \\
Smoothing & Savitzky-Golay, window 10, order 7, every 200 steps, averaged 1:1 with the current values; skipped below 10 frames \\
Loss & MSE of $\log_{10}(|M|+10^{-6})$ for mel power spectrograms $M$: FFT 512, hop 256, 64 bands, 60 to 7800 Hz \\
Noise & Same aspiration noise for both models at every step; fixed seed for the final render \\
Tract & 44 oral segments; nasal, frication, radiation and other optional features disabled \\
\bottomrule
\end{tabular}
\end{table}

Recognition is measured as in \citep{mo2024vocaltrax}. A frozen wav2vec 2.0 base model \citep{baevski2020wav2vec} with mean pooling and a linear head is trained only on 18,000 real recordings from 36 other speakers, selected on 6,000 recordings from 12 further speakers, and tested on the reconstructions. We average three head seeds and give 95\% intervals from a paired bootstrap over speakers (20,000 resamples). Pitch and voicing are measured with CREPE.

Timings use one NVIDIA A100 at batch size one in single precision, with the highest matrix-multiplication precision for full steps, and exclude compilation. The tract is timed on 1 to 60~s inputs made by repeating AudioMNIST recordings, and its backward pass returns the gradients for the diameters and the source.

\begin{table}[t]
\centering
\caption{Recognition by a classifier trained on real speech, and distances to the reference, on 600 AudioMNIST recordings. Brackets give 95\% intervals. Transfer renders our parameters with VocalTrax. $F_0$: RMS error in cents; V-F1: voicing F1 score.}
\label{tab:quality}
\setlength{\tabcolsep}{4pt}
\begin{tabular}{lcccc}
\toprule
 & Acc.\ (\%) $\uparrow$ & Log-mel $\downarrow$ & $F_0$ (ct) $\downarrow$ & V-F1 $\uparrow$ \\
\midrule
Real & 86.7 {\scriptsize [81.6, 91.1]} & & & \\
VocalTrax & 10.0 {\scriptsize [7.7, 12.3]} & \textbf{0.250} & 46.0 & 0.843 \\
Ours & \textbf{14.5} {\scriptsize [11.8, 17.0]} & 0.278 & \textbf{40.1} & 0.846 \\
Transfer & 13.8 {\scriptsize [10.9, 16.5]} & 0.310 & 41.1 & \textbf{0.849} \\
\bottomrule
\end{tabular}
\end{table}

\begin{table}[t]
\centering
\caption{Time per call in seconds on one A100 at batch size one. The tract columns add the forward and backward passes of the tract. The full step includes the shared LF source, the loss and the optimizer.}
\label{tab:speed}
\setlength{\tabcolsep}{4pt}
\begin{tabular}{rcccccc}
\toprule
 & \multicolumn{3}{c}{Tract, fwd + bwd} & \multicolumn{3}{c}{Full optimizer step} \\
\cmidrule(lr){2-4}\cmidrule(lr){5-7}
Audio & VocalTrax & Ours & Ratio & VocalTrax & Ours & Ratio \\
\midrule
1 s & 1.111 & 0.023 & 47.5 & 1.45 & 0.50 & 2.9 \\
8 s & 9.30 & 0.148 & 62.6 & 11.77 & 3.62 & 3.3 \\
16 s & 17.63 & 0.297 & 59.3 & 22.98 & 7.13 & 3.2 \\
30 s & 34.46 & 0.546 & 63.1 & 42.46 & 13.35 & 3.2 \\
60 s & 68.06 & 1.092 & 62.3 & 85.19 & 28.98 & 2.9 \\
\bottomrule
\end{tabular}
\end{table}

\subsection{Results}
\label{ssec:results}

\textbf{Recognition.} The classifier recognizes 86.7\% of the real recordings (Table~\ref{tab:quality}). VocalTrax reconstructions are recognized at chance level (10.0\%) and ours at 14.5\%, an improvement of 4.5 points (interval 2.2 to 6.6). The absolute rates are low, and the predictions for all synthetic conditions concentrate on a few digits, so these numbers indicate relative rather than practical intelligibility. The original VocalTrax study reports 73.5\% for real recordings and 20.0\% for VocalTrax reconstructions \citep{mo2024vocaltrax}, without specifying the backbone or the data split. Despite our efforts with different classifier hyperparameters, we were unable to reproduce these results. Since all conditions are scored by the same classifiers, the comparison between the two models is unaffected. On the log-mel objective our fits are 11.3\% worse (interval 10.1\% to 12.4\%), while the $F_0$ error drops from 46.0 to 40.1 cents.

\textbf{Transfer.} Rendering our parameters with the original VocalTrax changes the waveform by 5.7\% (median relative $\ell_2$ distance), and Fig.~\ref{fig:spec}(c) and (d) are nearly identical. Recognition remains at 13.8\%, 3.8 points above VocalTrax (interval 2.5 to 5.1). The log-mel MSE rises to 0.310, a paired increase of 0.032 over our renderer (interval 0.024 to 0.041). On the descriptors of Table~\ref{tab:mir}, the transferred audio stays close to ours, and its largest change is an 8.5\% higher log-STFT distance. Both are closer to the reference than VocalTrax in pitch, voicing, envelope and harmonic fraction, while VocalTrax is closer in spectral shape.

\begin{table}[t]
\centering
\caption{Further descriptors averaged over the 600 recordings. Pitch measures use the 581 recordings voiced in all conditions.}
\label{tab:mir}
\setlength{\tabcolsep}{4pt}
\begin{tabular}{lccc}
\toprule
Descriptor & VocalTrax & Ours & Transfer \\
\midrule
Log-STFT L1 $\downarrow$ & \textbf{1.566} & 1.579 & 1.714 \\
MFCC L2 $\downarrow$ & \textbf{7.19} & 7.80 & 7.74 \\
MFCC DTW cost $\downarrow$ & \textbf{6.35} & 6.66 & 6.46 \\
Centroid error (Hz) $\downarrow$ & \textbf{2587} & 2727 & 2813 \\
Contrast error (dB) $\downarrow$ & \textbf{5.27} & 5.83 & 5.35 \\
$\Delta$MFCC L1 $\downarrow$ & 0.589 & \textbf{0.572} & 0.576 \\
RMS envelope error ($10^{-2}$) $\downarrow$ & 1.50 & 1.46 & 1.46 \\
$F_0$ median error (ct) $\downarrow$ & 18.8 & \textbf{15.0} & \textbf{15.0} \\
$F_0$ gross errors (\%) $\downarrow$ & 0.47 & \textbf{0.21} & 0.24 \\
Harmonic fraction error $\downarrow$ & 0.176 & 0.173 & 0.173 \\
\bottomrule
\end{tabular}
\end{table}

\begin{figure}[t]
\centering
\includegraphics[width=0.6\linewidth]{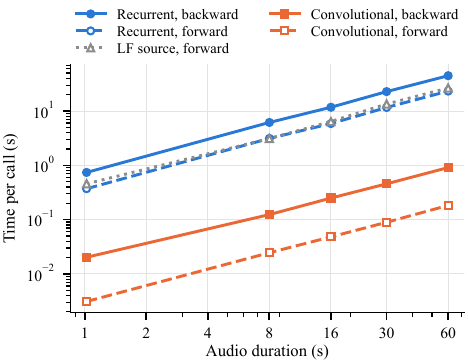}
\caption{Time per call of both tracts and the shared LF source, at batch size one on one A100.}
\label{fig:timing}
\end{figure}

\textbf{Speed.} The tract forward pass is 119 to 130 times faster than in VocalTrax and the backward pass 37 to 50 times faster, which gives the ratios of 47 to 63 in Table~\ref{tab:speed}, and both costs grow linearly with duration (Fig.~\ref{fig:timing}). A full optimizer step is about three times faster, and complete 1000-step fits of three AudioMNIST recordings took 12 to 18 minutes with VocalTrax and 3.8 to 5.8 minutes with ours. The rest of the step is spent mostly in the shared LF source, which we did not modify and which computes the pitch phase with a loop over samples (Fig.~\ref{fig:timing}). Subtracting its time gives an estimated step speed-up of 17 to 25 times.

\section{Discussion}
\label{sec:discussion}

\begin{figure}[t]
\centering
\includegraphics[width=0.6\linewidth]{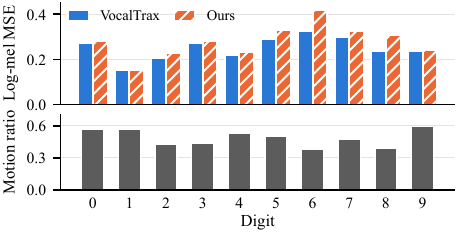}
\caption{Per digit (60 recordings each). Top: log-mel MSE. Bottom: frame-to-frame change of the four tract controls, ours divided by VocalTrax.}
\label{fig:digit}
\end{figure}

The better recognition despite a higher log-mel loss appears related to how much the fitted tract moves. Between consecutive frames, the four tract controls fitted by our model change about half as much as those of VocalTrax (ratios 0.46 to 0.52), while the tenseness changes 0.80 as much. For fixed parameters, the two renderers differ more when the reflection coefficients change faster (Spearman $\rho=0.45$), so the frozen-block model rewards fast tract motion less. Fig.~\ref{fig:digit} shows this per digit: ``six'' and ``eight'' lose the most motion (ratios 0.38 and 0.39) and have the largest loss gaps, while the fully voiced ``one'' and ``nine'' show almost none. The frozen blocks thus act as a temporal smoothness prior that costs spectral detail but discourages jittery trajectories, which the recurrent model may use to lower the loss without making the digit easier to recognize.

Our model also removes long products of Jacobians from the gradient. In the damped tract of VocalTrax these did not make the gradients vanish: with a waveform loss, the gradient norm scaled by the square root of the number of frames stayed constant from 0.5 to 60~s, consistent with the short memory in Table~\ref{tab:validation}. The benefit therefore lies in the cost and depth of the gradient computation. It does not imply a smoother loss, as our model showed much larger local curvature at initialization along a probed direction.

The speed-up concerns GPU execution. On a CPU the recurrent model needs only 0.15~s per step for a one-second clip, and on a GPU it reaches 514 clip updates per second at 1~s with large batches, against 73 for our model, which needs more memory (379 against 145~MiB per fit). Our model therefore suits single or long signals and GPU pipelines. Both models also inherit a source artifact. Unvoiced frames use a 1~Hz placeholder, which puts about 45\% of the output energy below 60~Hz, against 3.8\% in the recordings, and likely lowers absolute recognition. Finally, we use one dataset and one optimization seed.

\section{Future Work}
\label{sec:future}

Since $F_0$ is fixed during fitting, the pitch phase of the LF source can be computed once per clip, which should bring the end-to-end speed-up close to the estimated 17 to 25 times. Carrying the wave state across blocks could shorten the frozen tails, and a few recurrent steps after the convolutional fit may recover the lost spectral detail. We also plan to replace the unvoiced placeholder and to evaluate longer utterances with listening tests.

\bibliographystyle{plainnat}
\bibliography{refs,refs_local}

\end{document}